\documentclass[aps,prb,showpacs]{revtex4}

\usepackage{graphicx}
\usepackage{amsmath}

\graphicspath{{./IMAGES/}}

\begin{document}
\title{Searching for confined modes in graphene channels: the variable phase method}

\author{D. A. Stone}
\affiliation{School of Physics, University of Exeter, Stocker Road,
Exeter EX4 4QL, United Kingdom}

\author{C. A. Downing}
\affiliation{School of Physics, University of Exeter, Stocker Road,
Exeter EX4 4QL, United Kingdom}

\author{M. E. Portnoi}
\email[]{m.e.portnoi@exeter.ac.uk}
\affiliation{School of Physics,
University of Exeter, Stocker Road, Exeter EX4 4QL, United Kingdom}
\affiliation{International Institute of Physics, Av. Odilon Gomes de Lima,
1722, Capim Macio, CEP: 59078-400, Natal - RN, Brazil}

\date{August 24, 2012}

\begin{abstract}
Using the variable phase method, we reformulate the Dirac equation governing the charge carriers in graphene into a nonlinear first-order differential equation from which we can treat both confined-state problems in electron waveguides and above-barrier scattering problems for arbitrary-shaped potential barriers and wells, decaying at large distances. We show that this method agrees with a known analytic result for a hyperbolic secant potential and go on to investigate the nature of more experimentally realizable electron waveguides, showing that, when the Fermi energy is set at the Dirac point, truly confined states are supported in pristine graphene. In contrast to exponentially-decaying potentials, we discover that the threshold potential strength at which the first confined state appears is vanishingly small for potentials decaying at large distances as a power law, but nonetheless further confined states are formed when the strength and spread of the potential reach a certain threshold.
\end{abstract}


\maketitle

\section{\label{intro}Introduction}

After being constructed in the 1920s\cite{Courant} the variable phase method (VPM) was soon adapted to the theory of potential scattering by Morse and Allis\cite{MorseAllis} and has been expanded by numerous authors since as a method to solve scattering and confinement problems in quantum mechanics.\cite{Calogero63,Calogero,Taylor,Babikov} The method has proven to be robust in the non-relativistic case and has been applied to various physical problems, including an analysis of the statically screened Coulomb potential\cite{PortnoiVPM2D} and the ionization of the electron-hole plasma\cite{PortnoiIonisation,Ouerdane08} in conventional semiconductors. The method has also been used to find scattering lengths of colliding atoms\cite{Ouerdane03} and has been extended for use with non-local potentials.\cite{Kidun02,Kidun05} However, to the best of our knowledge, the method has never been applied to the two-dimensional modification of the Dirac-Weyl equation describing the low-energy spectrum of charge carriers in graphene.\cite{McClure,Dirac,CastroNeto} In this paper we study the presence of bound modes within a smooth, confining electrostatic potential well in graphene in a waveguide geometry.

It is generally accepted that purely electrostatic confinement of charge carriers in graphene is not possible due to the effect of Klein tunneling.\cite{Klein, Katsnelson, Reijnders} At finite energy, carriers inside a potential well couple to states outside the potential via same-energy states in the valence band. This is due to the conservation of chirality, which forbids backscattering for normally incident particles,\cite{Katsnelson} therefore the transmission probability is unity irrespective of the barrier height. However, for particles possessing a finite longitudinal wavevector, which is the situation for particles that are not normally incident on the barrier, the effect is diminished.\cite{Silvestrov} This suggests that a waveguide geometry, in which electrons with such a longitudinal wavevector propagate along the channel, may be appropriate for the creation of conducting channels in graphene sheets.

It has been previously demonstrated that the waveguide geometry for electrostatic potential wells does indeed lead to the lateral confinement of electrons in channels,\cite{SmoothWaveguides,Wu11,Williams11,Rozhkov} from which these confined modes provide an enhanced longitudinal conductivity. However, for energies away from the Dirac point the conductivity provided by the confined states is swamped by the sea of free carriers in the sample. This complicates the immediate application of graphene to digital electronics, which requires high on/off ratio current. It is therefore prudent to consider a graphene sheet close to the charge neutrality point, which can be easily achieved by adjusting the back-gate potential.

Exact analytical solutions for a smoothly-varying electrostatic confinement potential are only known for a somewhat unrealistic hyperbolic secant potential at zero energy,\cite{SmoothWaveguides,Exciton} and are also available at any energy for even less realistic square wells.\cite{Pereira}
The VPM allows one to probe the presence of confined states in any arbitrary potential channel as long as the potential decays faster than $1/x$ at large distances away from the central axis and is non-singular at finite distances. Therefore this method can act as a tool to determine the confined states in physically realizable electrostatic potentials. There is also continued interest in chiral tunneling and potential scattering problems in graphene.\cite{Reijnders, arxiv} The VPM provides an efficient numerical tool for treating above-barrier reflection from arbitrary-shaped decaying one-dimensional potentials at non-normal incidence.

The rest of this work is outlined as follows. After developing the formalism of the VPM in Sec.~\ref{formalism} we show that the method suggests a model top-gate structure should support zero-energy modes (Sec.~\ref{results}) and captures a hitherto unknown major distinction between exponentially-decaying and more realistic power-law decaying potentials. Namely, there is a threshold potential strength for the appearance of a confined zero-energy state in an exponentially-decaying potential, whereas there is no threshold for a power-law decaying potential. This observation is confirmed by general analysis in Sec.~\ref{threshold}. We discuss the impact of our results in Sec.~\ref{disc}.
Appendices~\ref{appendA}~and~\ref{appendB} contain detailed derivations of the analytic results for two exponentially-decaying potentials, which were used to validate the proposed variable-phase method.

\section{\label{formalism}Formalism}

In the low-energy approximation, charge carriers around a Dirac point in graphene are governed by the Dirac-Weyl Hamiltonian acting on a two-component wavefunction.\cite{Dirac}  In the presence of an external electrostatic potential $U(x)$ the Hamiltonian is given by\cite{Abergel}
\begin{equation}
\label{DiracEq}
\hat H = v_{\mathrm F} \left(\tau\sigma_x {\hat p_x} + \sigma_y {\hat p_y}\right) + U(x) + \tau\Delta\sigma_z,
\end{equation}
where $v_{\mathrm F}\approx c/300$ is the Fermi velocity of the charge carriers, $\sigma_{x,y,z}$ are the Pauli spin matrices and ${\hat p_{x,y}}=-i\hbar\partial_{x,y}$ are the linear momentum operator components. For completeness, and due to interest in the creation of confined states via band gaps in graphene, a mass term $\Delta$ is included in the final term in Eq.~\eqref{DiracEq}. Here $\tau=\pm1$ denotes whether the low-energy approximation is around the $K$ or $K'$ points. Since the potential varies only in $x$, the momentum operator $\mathit{\hat p}_y$ commutes with the Hamiltonian and so we may seek a solution in the form of the two-component wavefunction $\Psi(x,y) = \exp(iq_yy) [\psi_A(x),\psi_B(x)]^T$, where $q_y>0$ is the longitudinal wave vector along the waveguide and $\psi_{A(B)}$ refer to the wave-function components associated with the inequivalent $A(B)$ sublattices in graphene. These wave-function components satisfy the coupled first-order differential equations
\begin{align}
\label{coupledpsi}
 \begin{split}
  \left(\tau\frac{d}{dx} +q_y \right) \psi_B &= (\varepsilon-V(x) - \tau\delta)i\psi_A, \\
  \left(\tau\frac{d}{dx} -q_y \right) \psi_A &= (\varepsilon-V(x) + \tau\delta)i\psi_B,
 \end{split}
\end{align}
where $\varepsilon=E/\hbar v_\mathrm F$, $E$ is the energy eigenvalue, $V(x)=U(x)/\hbar v_\mathrm F$ which we assume to be rapidly vanishing as $x\to\pm\infty$ and $\delta=\Delta/\hbar v_\mathrm F$. Notably, the sign of the potential is not important in considerations of zero-energy states in gapless graphene, as can be seen by reducing the system of Eqs.~\eqref{coupledpsi} to a second-order differential equation (see Appendix~\ref{appendA}). Incidentally, there has been a recent rise in interest in zero-energy states in graphene, due to both a curiosity in edge states\cite{ZhangChang, ZareniaChaves, GrujicZarenia} and the possibility of observing Majorana zero-modes in this system.\cite{Sodano, Jackiw}

Following the work of Babikov,\cite{Babikov} we write a wavefunction component as a superposition of transmitted and reflected waves
\begin{align}
\label{WFtransref}
\psi_A(x) = C(x) \left( e^{iq_xx} + D(x) e^{-iq_xx} \right),
\end{align}
where $C(x)$ and $D(x)$ are known as the transmission functions and reflection functions respectively, and propagation is taken to be in the positive $x$ direction. In addition to the introduction of the transmission and reflection functions, we will impose a condition on the first derivative of the upper wavefunction component, as we are free to do to completely define $C(x)$ and $D(x)$.

Our choice of wavefunction in Eq.~\eqref{WFtransref} is a modification of the free-particle solution which has the energy spectrum $\varepsilon=\pm\left(q_x^2+q_y^2+\delta^2\right)^{1/2}$, which includes a band gap of $2\delta$.\cite{Abergel} The amplitude functions $C(x)$ and $D(x)$ have a natural interpretation if one considers the potential to possess cutoffs at a point $x_1$ to the left of the origin and $x_2$ to the right of the origin. Then the barrier exists only in the region $[x_1,x_2]$; outside of this region the amplitude functions are not expected to be position dependent, thus $C(x_2)$ is simply the transmission amplitude and $C(x_1)D(x_1)$ is the reflection amplitude.\cite{Kidun05} For a fast-decaying potential the amplitude functions tend to these position-independent values away from the center. Since the considered problem contains only a single electron and the electrostatic potential well, one may assert boundary conditions on the amplitude functions. The ingoing amplitude is unity, $C(x\to-\infty)=1$, and the reflection on the other side of the boundary is zero, $D(x\to+\infty)=0$. The reflection coefficient\cite{Babikov} can then be simply taken to be $R=|D(x\to-\infty)|^2$.

In implementing the VPM we make the following aforementioned ansatz on the first derivative of the upper wavefunction component
\begin{equation}
\label{WFansatz}
\frac{d}{dx}\psi_A(x) = C(x) \left( \frac{d}{dx}e^{iq_xx} + D(x) \frac{d}{dx}e^{-iq_xx} \right).
\end{equation}
The ansatz indeed agrees with direct differentiation of the wavefunction in the limit of large $x$, when the amplitudes become constants. However, this choice of ansatz is expedient because it always allows the transmission function to be eliminated later on (resulting in a differential equation purely in terms of the reflection function). Equating the right-hand side of Eq.~\eqref{WFansatz} with the full derivative of the wavefunction yields the following useful relation for the transmission function
\begin{align}
\frac{d C(x)}{d x} &= -C(x) \frac{d D(x)}{d x} \frac{ e^{-iq_xx} }{e^{iq_xx} + D(x)e^{-iq_xx}}.
\end{align}
Upon substituting Eqs.~(\ref{WFtransref}) and (\ref{WFansatz}) into the second of Eqs.~\eqref{coupledpsi} we find the lower component of the wavefunction to be
\begin{align}
\psi_B(x) = \frac{C(x)}{\varepsilon-V(x)+\tau\delta} \left[ (\tau q_x+iq_y) e^{iq_xx} + (iq_y-\tau q_x)D(x)e^{-iq_xx}\right].
\end{align}
This is exactly the form for $\psi_B$ that one would expect from a flat potential profile, since it reflects the pseudospinor nature of the components. We may now employ the complete wavefunction and the first of the coupled Eqs.~\eqref{coupledpsi}, and use Eq.~\eqref{WFansatz} to eliminate the transmission function, resulting in a differential equation only containing the reflection function:
\begin{align}
\label{BODE}
 \frac{d D(x)}{d x} = -\frac{f_A(x)}{2q_x} \left[- \tau\frac{dV(x)}{dx} f_B(x) + i\left( \left(\varepsilon-V(x)\right)^2 - \varepsilon^2 \right) f_A(x) \right],
\end{align}
where the auxiliary functions $f_{A,B}(x) = \psi_{A,B}(x)/C(x)$. This is a nonlinear, first-order differential equation of the Riccati  type and does not permit an analytical solution for an arbitrary potential. Numerical attempts to solve this equation will break down if $\varepsilon-V(x)+\tau\delta=0$, which would occur for a potential in which the Fermi energy crosses both electron-like and hole-like regions. For real wavevectors, Eq.~\eqref{BODE} describes scattering problems for Dirac fermions; reflection and transmission amplitudes can easily be found and the signature of resonant tunneling through the barrier is indicated by zeros in the reflection amplitude at certain values of energy and the parameters of the potential. For imaginary wavevectors, Eq.~\eqref{BODE} allows one to tackle confinement problems for Dirac fermions, including analysis of bound and quasi-bound states, which are of special interest due to the phenomenon of Klein tunneling.\cite{Klein,Katsnelson, Reijnders}

We are interested in probing the presence of bound modes in the potential at zero energy for the massless Dirac fermions of pristine graphene, thus we must set $\delta=0$ and $q_x=\pm iq_y$. In the massless case the charge carriers exhibit valley symmetry, allowing us to confine ourselves to only one $K$ point by setting $\tau=1$ without loss of generality. In making the choice of complex wavevectors, the interpretation of the reflection probability from the potential barrier is that bound-state resonances are indicated by divergences (peaks) in $R$.\cite{Feynman} As a result, solving the differential equation numerically becomes almost intractable. These divergences are overcome by applying a convenient regularization of the form
\begin{align}
\label{regularisation}
D(x) = \tan[\theta(x)],
\end{align}
and rewriting the differential equation in terms of the phase function $\theta(x)$. Divergences in $D(x)$ are mapped onto $\theta(x)=\pm \pi/2$, thus the numerical method becomes manageable. This is familiar from the Levinson theorem applied to the one-dimensional Dirac problem,\cite{LinLevinsonDirac2D,MaDong06} in that the appearance of a new bound state corresponds to an accumulated scattering phase shift of $\pi$. Taking $q_x=iq_y$, the first-order differential equation \eqref{BODE} in terms of $\theta(x)$ for $\varepsilon=0$ reduces to
\begin{align}
\label{thetaODE}
 \begin{split}
  \frac{d\theta}{dx} &= - f(x) \left[ \frac{1}{V(x)}\frac{dV(x)}{dx}\cos[\theta(x)]e^{-q_yx} + \frac{V(x)^2}{2q_y} f(x) \right], \\
  f(x) &= \cos[\theta(x)]e^{-q_yx} + \sin[\theta(x)]e^{q_yx}.
 \end{split}
\end{align}
This differential equation can be solved by employing the initial condition $\theta(x\to+\infty)=0$, which arises due to
zero refection amplitude in the region far behind the potential, and solving as an initial value problem. Eq.~\eqref{thetaODE} is a stiff differential
equation so it requires implicit multi-step methods in order to accurately proceed to $x\to-\infty$. The reflection coefficient $R=|D(x\to-\infty)|^2$ can then be found.\cite{Babikov}

\section{\label{results}Results}

\begin{figure}
 \includegraphics[width=0.6\textwidth]{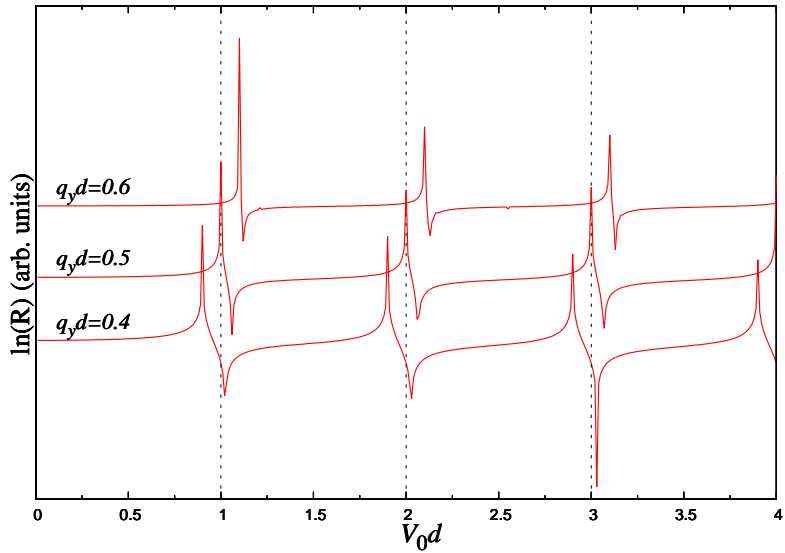}
 \caption{Reflection coefficient vs the potential strength for zero-energy charge carriers incident on a hyperbolic secant potential of unit width, for several values of propagating wave vector along the waveguide. Confined states are indicated by the divergences (peaks). The plots are displaced vertically from each other for clarity.}
 \label{fig:triplething}
\end{figure}
We can now implement the VPM for massless Dirac fermions into physical problems. It is instructive to first demonstrate that this numerical method can reproduce analytical results from the literature. An exact result has been stated for an electron waveguide defined by the model potential $V(x)=-V_0/\cosh(x/d)$ in Ref.~[\onlinecite{SmoothWaveguides}]. However, since the derivation of this result\cite{HartmannBook} has been left out of the widely available literature, we outline a procedure to obtain the exact solution in Appendix~\ref{appendA}. In this case, the condition for zero-energy confined states to be supported is $|V_0d|=q_yd+(n+\frac 1 2)$, where $n$ is a non-negative integer and the threshold for the appearance of the first mode is $|V_0d|>\frac 1 2$. For example, if we choose a potential with unit characteristic width ($d=1$) and consider modes with propagating wavevector $q_y=\frac1 2$ we expect zero-energy modes to appear at integer values of potential strength $|V_0|$. In Fig.~\ref{fig:triplething}, we plot the reflection coefficient $R$ against potential strength for three values of $q_y$, including $q_y=\frac1 2$. The positions of the positive spikes in the reflection coefficient correspond to bound states, verifying the analytically-derived condition.

\begin{figure}[tbp]
 \begin{centering}
  \setlength{\unitlength}{1.0\textwidth}
  \includegraphics[width=0.35\textwidth]{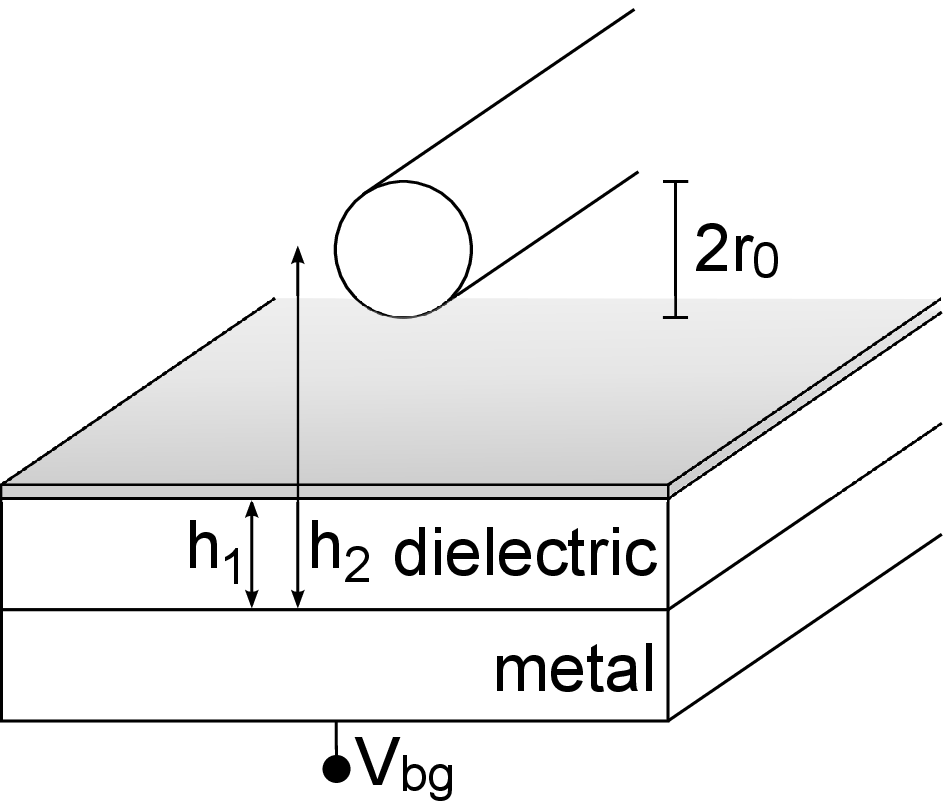}
  \put(-0.35,0.275)  { \makebox(0,0)[c] {(a)} } \qquad
  \includegraphics[width=0.45\textwidth]{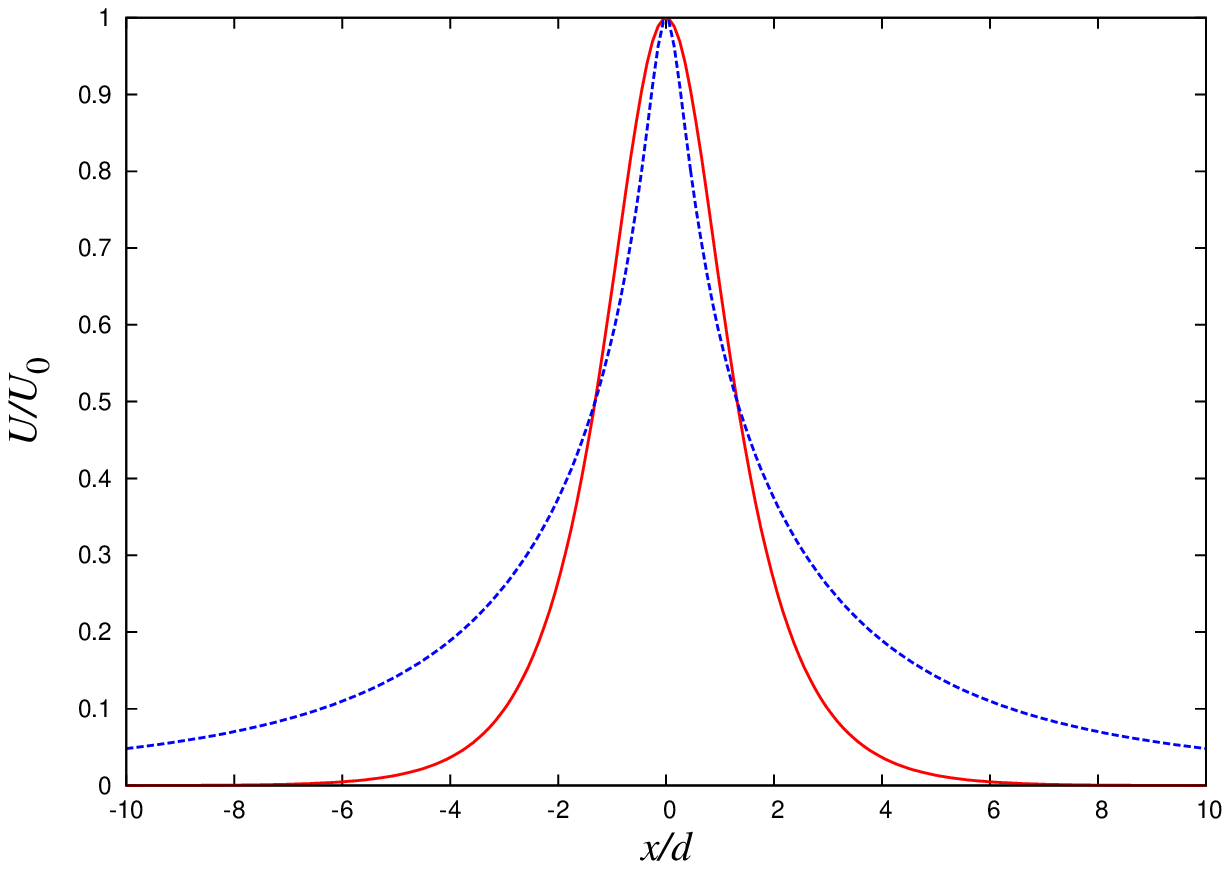}
  \put(-0.38,0.275) { \makebox(0,0)[c] {(b)} }
 \end{centering}
 \caption{(a) A schematic diagram of the experiment being considered. The top-gate is charged and produces an electrostatic potential in the graphene plane. The top-gate is treated as a wire of radius $r_0$ suspended above the graphene plane. The graphene layer is maintained at charge neutrality by charging the back-gate to $V_{bg}$. (b) Comparison of the realistic top-gate structure (dotted blue line) to the fitted hyperbolic secant model (solid red line), for $h_1/h_2=0.9$.}
 \label{fig:topgate}
\end{figure}
The value of the proposed method is that it can be employed with any desired potential, including those not analytically solvable. Let us then consider a more physically realizable electrostatic potential, namely the potential created by a top-gate structure as in Fig.~\ref{fig:topgate}(a). This structure can be formed by the deposition of a thin metallic strip onto an insulating layer of material on the graphene sheet. The insulating layer may be removed, for example, by solvents, producing a so-called `air bridge'.\cite{Savchenko, Lau08} A model top-gate potential can be obtained, to first order, from the well-known method of image charges\cite{SmoothWaveguides}
\begin{equation}
\label{TopGate}
U_t(x) = \frac{e\Upsilon}{2}\ln\left(\frac{x^2 + (h_2-h_1)^2}{x^2 + (h_2+h_1)^2}\right),
\end{equation}
where $h_1$ and $h_2$ are the distances between the metallic back gate and the graphene sheet and charged top gate, respectively, and $\Upsilon$ is matched to the top-gate voltage by
\begin{equation}
\Upsilon = \frac{ V_{tg} }{ \ln \left( \dfrac{2h_2 - r_0}{r_0} \right) },
\end{equation}
where $r_0$ is the characteristic width of the top gate. For simplicity, we assume the dielectric constant of the dielectric layer between the graphene and metallic back gate to be equal to unity, which corresponds to the experimentally attainable case of suspended graphene.\cite{suspended12} The presence of a dielectric can be easily accounted for, again with the method of images;\cite{Jackson} however, it does not noticeably change the functional dependence of the potential, and especially its long-range asymptotics. To compare the top gate and hyperbolic secant waveguides we use a simple fitting procedure: fixing the potential maximum at the origin and the half width at half-maximum.\cite{SmoothWaveguides} This yields the following expressions for $V_0$, the absolute value of the potential at $x=0$, and $d$ in terms of top-gate parameters, for which the hyperbolic waveguide potential will best match that produced by the realistic top gate,
\begin{align}
V_0 = \frac{ e\Upsilon }{ \hbar v_F  } \ln \left( \dfrac{h_2 + h_1}{h_2 - h_1} \right), \quad
d   = \frac{ \sqrt {h_2^2 - h_1^2 }}{ \operatorname{arccosh} (2)  }.
\end{align}
%
For unit characteristic width ($d=1$) of the hyperbolic waveguide we show an example of this fitting procedure in Fig.~\ref{fig:topgate}(b), where we are using feasible experimental parameters:\cite{Savchenko} $h_1\approx 2.7d$, $h_2\approx 3.0d$, and $V_0\approx 2.94e\Upsilon/\hbar v_\mathrm F$. We can see that close to the origin the potentials fit well, since both functions are approximately parabolic. At large distances the behavior is very different, however; the secant hyperbolic waveguide is exponentially suppressed, whereas Eq.~\eqref{TopGate} falls as $1/x^2$. This difference motivates this work because different behavior would be expected for states which possess quasiclassical turning points at significantly different distances from the origin. There is also a fundamental mathematical difference between the two confinement potentials under consideration; namely, the different behavior of the logarithmic derivative of the potential, which features in Eq.~\eqref{thetaODE}. In general the logarithmic derivative term will reach zero asymptotically for a power-law decay, whereas it will tend to a constant value for any exponential damping. Since an exponentially damped electrostatic potential is not realistic, we should be cautious when drawing general conclusions from such a model function.

\begin{figure}[ht]
 \begin{centering}
  \setlength{\unitlength}{1.0\textwidth}
  \includegraphics[width=0.4\textwidth]{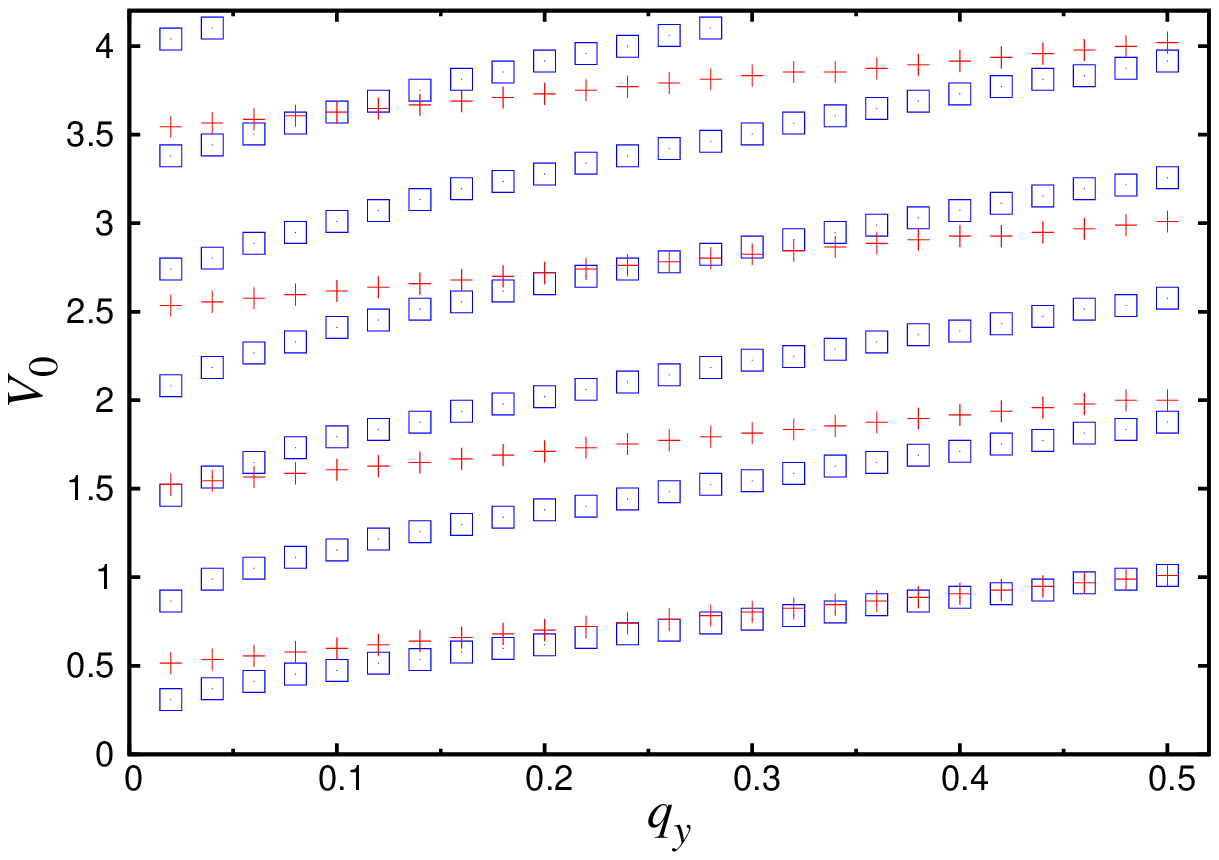}
  \includegraphics[width=0.4\textwidth]{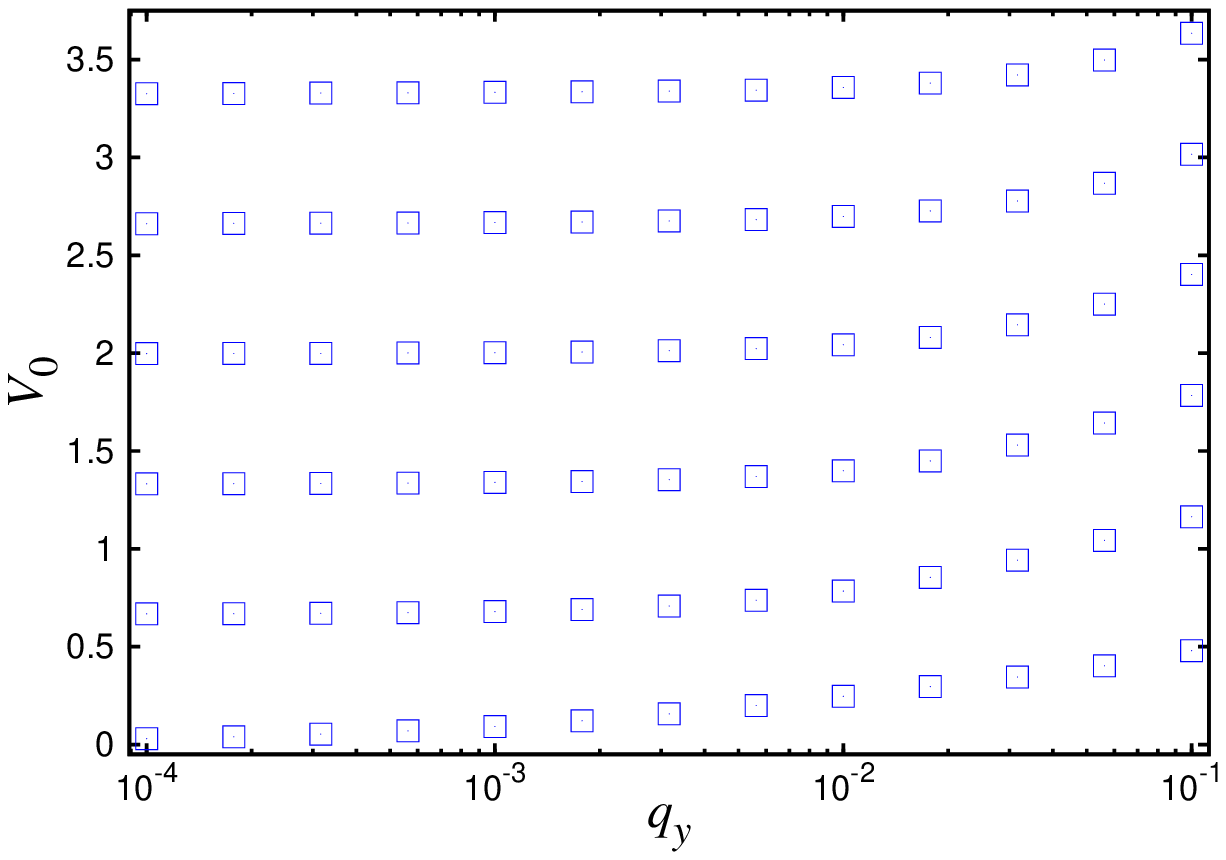}
 \end{centering}
 \caption{(Color online) The appearances of bound states for the top-gate potential and the fitted exactly-solvable hyperbolic secant waveguide, with $h_1/h_2=0.9$ and $d=1$. Boxes and crosses indicate propagating modes for the model and realistic top-gate potential correspondingly. The left panel shows the large-scale plot for $q_y\in[0,0.5]$. The right panel contains a logarithmic plot for small $q_y$ for the realistic potential only, indicating that the threshold for the power-decaying top-gate potential is vanishingly small.}
 \label{fig:scatterpeaks}
\end{figure}

We compare the results of the analytic solution\cite{SmoothWaveguides} and the more realistic potential in Fig.~\ref{fig:scatterpeaks}(a). As expected, in both cases new propagating states appear fairly regularly with increasing potential depth. The appearance of bound states for the hyperbolic secant waveguide follows a linear relation between $V_0$ and $q_y$ and new states appear equally spaced according to the analytical relationship. The top-gate also produces new states following a roughly linear relationship for large propagating wavevectors. Such a similarity is to be expected when one notes that an electron possessing a large propagating wavevector will have quasiclassical turning points close to the origin, where the potentials possess similar functional forms.

The appearance of confined modes for the realistic top-gate potential does not follow a linear relationship for lower $q_y$; instead as $q_y$ decreases, the values of $V_0$, at which the first states appear, fall off dramatically. In fact, contrary to the hyperbolic secant waveguide (and indeed the exponentially decaying cusp waveguide which also yields an exact solution, see Appendix~\ref{appendB}), which possesses a finite threshold value, for the realistic top-gate waveguide case we find a complete absence of a threshold, as shown in Fig.~\ref{fig:scatterpeaks}(b). This could be attributed to the logarithmic derivative term, which acts drastically different in the two cases (as we shall see in Sec.~\ref{threshold}) such that only potentials with exponential tails have a threshold potential strength. This should lead to an enhanced conductivity in a graphene sheet for waveguide-type devices even for vanishingly weak potential channels. Interestingly, the zero threshold for a power-law decay of the confining potential does not apply universally to charge carriers in graphene. It was shown recently in the radial geometry that electrostatic potentials without exponentially decaying tails may indeed possess a threshold for the appearance of the first zero-energy state.\cite{Vortices, Calvo}
%

\section{\label{threshold}General analysis of power-law decaying potentials}

It has been stated in Ref.~[\onlinecite{SmoothWaveguides}] (see also Appendix~\ref{appendA}) and shown numerically in Sec.~\ref{results} that for the hyperbolic secant waveguide there exists a threshold potential strength, below which there are no bound states. Here we show that this property does not extend to potentials that decay as a power law, such as the considered model top-gate potential Eq.~\eqref{TopGate}, due to the logarithmic derivative of the potential tending towards zero asymptotically.

First, let us consider the case of a general exponentially-decaying potential $V(x)=V_0/\cosh^t(x/d)$ with parameter $t>0$. As the logarithmic derivative will always tend to a constant asymptotically, in the limit $x\to\pm\infty$, it follows from Eqs.~\eqref{coupledpsi} that
\begin{equation}
\label{Decoupled4}
\frac{d^2}{dx^2}\psi_B(x) +\frac{t}{d}\frac{d}{dx}\psi_B(x) + \left(\frac{tq_y}{d} - q_y^2\right) \psi_B(x) = 0,
\end{equation}
and the equation for $\psi_A(x)$ is obtained upon replacing $q_y$ with $-q_y$ in Eq.~\eqref{Decoupled4}. Hence the normalizable solutions describing the first emergent state of both wavefunction components in the region $x>0$ are
\begin{equation}
\label{positive2}
\psi_B(x) \sim e^{-q_{y}x}, \qquad \psi_A(x) \sim e^{-(q_{y}+t/d)x}
\end{equation}
and accordingly in the region $x<0$
\begin{equation}
\label{negative2}
\psi_B(x) \sim e^{(q_{y}-t/d)x}, \qquad \psi_A(x) \sim e^{q_{y}x}.
\end{equation}
Thus, we arrive at the condition $q_{y}d > t$ to ensure a non-trivial normalizable solution in the limit of large negative $x$ and so it is apparent in this scenario we do indeed have a threshold, as we expect from our numerical solutions in Sec.~\ref{results}.
A particular case of the hyperbolic secant potential is considered in Appendix~\ref{appendA}.
The presence of a threshold in the product of potential strength and spread for the appearance of the first confined mode can also be shown explicitly for another exponentially-decaying potential supporting analytic zero-energy solutions which is considered in Appendix~\ref{appendB}.

Now let us turn our attention to the equation governing the behavior of the lower wavefunction component $\psi_B(x)$ in the potential $V(x)=V_0/(1+\frac{x^2}{d^2})^{p/2}$, where the decay is characterized by $p\ge1$. In the limit $x\to\pm\infty$ Eqs.~\eqref{coupledpsi} yield
\begin{equation}
\label{Decoupled3}
\frac{d^2}{dx^2}\psi_B(x) +\frac{p}{x}\frac{d}{dx}\psi_B(x) + \left(\frac{pq_y}{x} - q_y^2\right) \psi_B(x) = 0,
\end{equation}
and a similar equation for $\psi_A(x)$ can be written by changing $q_y$ to $-q_y$ in Eq.~\eqref{Decoupled3}. Both of these equations can easily be reduced to a confluent hypergeometric differential equation, known as Kummer's differential equation.\cite{Abramowitz} The square-integrable solutions describing the first emergent state for both wavefunction components when $x>0$ are\cite{Confluent}
\begin{equation}
\label{positive1}
\psi_B(x) \sim e^{-q_{y}x}, \qquad \psi_A(x) \sim e^{-q_{y}x}~\mathcal{U}(p,p, 2 q_{y} x),
\end{equation}
where $\mathcal{U}$ denotes the confluent hypergeometric function of the second kind, or Tricomi function. The corresponding square-normalizable solutions for $x<0$ are\cite{Confluent}
\begin{equation}
\label{negative1}
\psi_B(x) \sim e^{q_{y}x}~\mathcal{U}(p,p, -2 q_{y} x), \qquad \psi_A(x) \sim e^{q_{y}x}.
\end{equation}
The terms including the Tricomi functions decay quickly at large $x$ (in the region considered) for all decay strengths $p$ and propagating wave vectors $q_{y}$, as can be seen via a series expansion at infinity.\cite{Abramowitz} The functional forms of the wave-function components at large distances, Eqs.~(\ref{positive1},\,\ref{negative1}), indicate that no conditions are required to be imposed upon the longitudinal wave
vector $q_y$ and hence there is no threshold of potential strength at which bound states first appear.

\section{\label{disc}Summary}

The variable phase method is elegant because it allows one to solve the Schr\"{o}dinger equation directly for physical quantities, such as the reflection coefficient or the scattering phase, rather than needing to extract these properties from the wave function.\cite{Kidun02} 
We have shown that this method can be extended to the Dirac-Weyl equation. 
We expect this method to become broadly used for modeling top-gate devices in the considered waveguide geometry, as well as for calculating barrier transparency at non-normal incidence, by computing reflection and transmission coefficients. 
The VPM also offers significant practical advantages: whilst it is always possible to reduce two coupled first-order differential equations into a decoupled second-order equation, the VPM allows one to decouple into a single first-order equation, from which a useful physical property can be immediately obtained. Of course, the nonlinear nature of the equation is of little importance when integrating numerically.

In conclusion, we have modified the variable-phase method so that it can be used to study quasi-one-dimensional problems, such as electron channels, for the quasi-relativistic charge carriers of graphene. The method has been validated by reproducing the exact result for a hyperbolic secant waveguide. We have gone on to find, for a model top-gate electrostatic potential with experimentally obtainable parameters, that there is no potential strength threshold for bound state emergence, so the first confined mode exists for an arbitrarily weak power-decaying potential. With increasing potential strength further confined states appear in steps of applied top-gate voltage of the order of tens of millivolts.

In fact, we have found that while there is always a threshold potential strength for the appearance of the first confined zero-energy state in channels defined by a potential with exponential tails, there is no such threshold in potentials decaying as a power law. This suggests that both the square waveguide model and indeed models with exponential tails are somewhat unsatisfactory, because they miss an important physical property necessarily
present in more realistic top-gate defined potentials. The presence of zero-energy modes for all realistic power-law decaying waveguide potentials 
can be related to the well-known problem of non-vanishing conductivity in graphene for the case of smooth disorder. 
Our results provide an additional argument in favor of the so-called resistor network model for minimum conductivity.\cite{Cheianov}

Electrostatically-defined waveguides are free from apparent shortcomings of graphene nanoribbons such as strong back-scattering produced by edges and technological difficulties in controlling nanoribbon parameters. Moreover a top-gate controlled potential prevents edge-related scattering and can serve as an analog of an electrostatically-defined carbon nanotube with an enhanced mean free path. We expect truly confined states to be observable in charge neutral pristine graphene without the need of magnetic fields, strain engineering or the introduction of band-gaps, a potential way forward in the pursuit of graphene devices aimed at digital manipulation of current.

\section*{Acknowledgments}
We thank R. R. Hartmann, N. J. Robinson and H. Ouerdane for fruitful discussions and A. M. Aleexev for a critical reading of the manuscript. This work was supported by the Millhayes Foundation (DAS), the EPSRC (CAD), the EU FP7 ITN Spinoptronics (Grant No. FP7-237252), FP7 IRSES projects SPINMET (Grant No. FP7-246784), TerACaN (Grant No. FP7-230778) and ROBOCON (Grant No. FP7-230832).

\begin{appendix}
\section{\label{appendA}Exact solution of the hyperbolic secant waveguide}
This appendix illustrates a method of solving the massless Dirac-Weyl equation for the case of a hyperbolic secant electron waveguide at zero-energy, following previous work in Ref.~[\onlinecite{HartmannBook}]. The wavefunctions and threshold conditions for fully-confined states in this potential were stated in Ref.~[\onlinecite{SmoothWaveguides}] without derivation. Since the aforementioned threshold conditions were used as the major benchmark for checking validity of the formulated VPM, we found it appropriate to provide here the analytic solution of the hyperbolic secant problem.

Upon setting $E=0$ and decoupling Eqs.~\eqref{coupledpsi} into a second-order differential equation for a single wavefunction component $\psi_B(x)$ only,  we obtain
\begin{equation}
\label{Decoupled}
\frac{d^2}{dx^2}\psi_B(x) -\frac{1}{V(x)}\frac{dV(x)}{dx}\frac{d}{dx}\psi_B(x) + \left(V(x)^2 - \frac{q_y}{V(x)}\frac{dV(x)}{dx} - q_y^2\right) \psi_B(x) = 0.
\end{equation}
Considering the one-dimensional potential $V(x)=-V_0/\cosh(x/d)$ and parameters $q_y>0$, $d>0$ and making the change of variable $\xi=\tanh(x/d)$ yields
\begin{equation}
\label{Known}
(1-\xi^2)^2\frac{d^2}{d\xi^2}\psi_B(\xi) -2\xi(1-\xi^2)\frac{d}{d\xi}\psi_B(\xi) + \left[ \omega^2(1-\xi^2) +\xi\delta + \delta^2 \right] \psi_B(\xi) = 0.
\end{equation}
where $\omega=|V_0d|$ and $\delta=q_yd$. This is a known differential equation which should be reduced using the following form of the solution\cite{Kamke}
\begin{equation}
\psi_B(\xi) = (\xi+1)^p(\xi-1)^q \eta \left[ \frac 1 2 (\xi+1) \right],
\end{equation}
where the function $\eta(\kappa)$ is to be found and $p$ and $q$ are subject to the conditions
\begin{align}
\label{pandqconditions}
\nonumber
4q(q-1)+2q+a+b+c=0, \qquad
(p-q)[2(p+q)-1]=c,
\end{align}
\begin{align}
\nonumber
a=-\omega^2, ~b=\delta, ~c=\omega^2-\delta^2.
\end{align}
Satisfying the conditions for $p$ and $q$ and making a further change of variable $\kappa = (\xi+1)/2$ leads to a differential equation in the well-known Gauss hypergeometric form
\begin{equation}
\label{Gaussform}
\kappa(\kappa-1)\frac{d^2}{d\kappa^2}\eta(\kappa) + \left[(2p+2q-1)\kappa - \left(2p+\frac 1 2\right)\right] \frac{d}{d\kappa}\eta(\kappa) + \left[ (p+q)^2 + a \right] \eta(\kappa) = 0,
\end{equation}
from which one can write down the unnormalized wavefunction solution as
\begin{equation}
\psi_B(\xi) = (1+\xi)^p(1-\xi)^q \, _2F_1\left( p+q+\omega, p+q-\omega; 2p+\frac 1 2; \frac{1-\xi}{2} \right),
\end{equation}
where $p=\frac{\omega-n}{2}-\frac 1 4$, $q=\frac{\omega-n}{2} + \frac 1 4$ and  $n$ is a non-negative integer. To avoid a singularity at $\xi=\pm1$ one obtains the condition that $\omega-n > \frac 1 2$. This puts an upper limit on $n$, such that for a channel of given parameters there may exist only a finite number of distinct propagating states. Termination of the hypergeometric series via $|V_0d|-q_yd=n+\frac{1}{2}$ gives the condition for which confined states are supported. The upper wavefunction component, $\psi_A$, can be obtained from the coupled Eqs.~\eqref{coupledpsi} and has a similar form to $\psi_B$.

\section{\label{appendB}Exact solution of the exponentially decaying cusp waveguide}
This appendix details the solution of the massless Dirac equation at zero-energy for an exponentially decaying cusp electrostatic potential\cite{Villalba} defined by $V(x)=V_0 \exp(-|x|/d)$, with parameters $q_y>0$, $d>0$.
The aim of this appendix is to confirm our general statement on the presence of threshold in the product of potential strength and its spatial extent for the appearance of the first confined state in exponentially-decaying potential. Besides, to our knowledge, a solution of the two-dimensional Dirac-Weyl equation for this potential has not been reported in the literature before.

The second-order differential equation to be solved for wavefunction component $\psi_B(x)$ follows from Eq.~\eqref{Decoupled}
\begin{equation}
\label{Decoupled2}
\frac{d^2}{dx^2}\psi_B(x) + \operatorname{sgn}(x)\frac{1}{d}\frac{d}{dx}\psi_B(x) + \left(V_0^2 e^{-2|x|/d} + \operatorname{sgn}(x)~\frac{q_y}{d} - q_y^2\right) \psi_B(x) = 0,
\end{equation}
where $\operatorname{sgn}(x)$ is the signum function of the coordinate $x$. This equation is of a standard form and yields the following solution in terms of Bessel functions of the first kind\cite{Kamke}
\begin{equation}
\label{SolutionExp}
\psi_B(x) = C_1e^{-x/2d}~J_{q_y d - 1/2} \left( |V_0 d|~e^{-x/d} \right) + C_2e^{x/2d}~J_{q_y d + 1/2} \left( |V_0 d|~e^{x/d} \right),
\end{equation}
where the first term in the solution is present only for $x\ge0$ and the second term only for $x\le0$. Again, the upper wavefunction component $\psi_A$ can be found via the coupled Eqs.~\eqref{coupledpsi}. Continuity of both wavefunction components at $x=0$ requires $C_1=C_2$ and yields the constraint
\begin{equation}
\label{Condition2}
J_{q_y d - 1/2} \left( |V_0 d| \right) = \pm J_{q_y d + 1/2} \left( |V_0 d| \right).
\end{equation}
Equation~\eqref{Condition2} is a transcendental equation, and graphical or numerical solutions display a threshold value of $|V_0d| = 0.785...$ for which the potential can support its first bound state, and indeed the values of potential strength at which further bound modes appear.

\end{appendix}


\begin{thebibliography}{100}
\bibitem{Confluent}The solutions that would involve $\Phi$, the confluent hypergeometric function of the first kind, or Kummer function, have been simplified using the identity $\Phi(a,a,z)= e^{z}$.
\end{thebibliography}
\end{document}